\documentclass[preprint,floats,aps,12pt,superscriptaddress,nofootinbib]{revtex4}
\usepackage{natbib} 
\usepackage{times} 
\usepackage{amssymb,amsmath}
\usepackage{prelim2e}\usepackage[none,bottom]{draftcopy}
\draftcopyName{preprint / }{1.2} %ADAPT TEXT
\ifx\pdfoutput\undefined
\usepackage[usenames,dvips]{color}
\usepackage{graphicx}     % Include figure files
\else
\usepackage[usenames,dvipsnames]{color}
\usepackage{epstopdf}
\usepackage[pdftex]{graphicx}
\usepackage[pdftex]{hyperref}
\hypersetup{a4paper,
    pdftitle={Manuscript on XXX} %ADAPT TEXT 
    pdfauthor={et al. and Uwe Thiele},  %ADAPT TEXT 
pdfproducer={lateX},
pdfview=FitV,       % FitH
pdfstartview=FitB,
linkcolor=blue,     % links to same page
citecolor=blue,     % citations
pagecolor=blue,     % links to same page
urlcolor=red,      % links to URLs
breaklinks=true,    % links may be split onto 2 lines
colorlinks=true,
citebordercolor=0 0 0,  % color for \cite
filebordercolor=0 0 0,
linkbordercolor=0 0 0,
menubordercolor=0 0 0,
pagebordercolor=0 0 0,
urlbordercolor=0 0 0,
pdfhighlight=/I,
pdfborder=0 0 0,   % no box around links
backref=false,
pagebackref=false,
bookmarks=true,
bookmarksopen=true,
bookmarksnumbered=true
}
\fi
\ifx\pdfoutput\undefined
\DeclareGraphicsExtensions{.eps}
\else
\DeclareGraphicsExtensions{.jpg, .pdf, .tif, .png}
\fi
\graphicspath{{.},{./Fig/}} %$ %ADAPT DIRECTORY
\usepackage[usenames,dvipsnames]{color}
\usepackage[normalem]{ulem}
\begin{document}
%
%----------------------------------------------------------------%
\title{The collective behaviour of ensembles of condensing liquid drops on heterogeneous inclined substrates\protect\footnote{In contrast to the published version, here we have directly included the Supplementary Material as section~\ref{sec:supp} after the Conclusion. Furthermore, here all main sections are numbered.}}
%\shorttitle{Drop ensembles condensing onto heterogeneous inclined substrates}

\author{Sebastian Engelnkemper}
\affiliation{Institut f\"ur Theoretische Physik, Westf\"alische Wilhelms-Universit\"at M\"unster, Wilhelm Klemm Str.\ 9, 48149 M\"unster, Germany}
\author{Uwe Thiele}
\affiliation{Institut f\"ur Theoretische Physik, Westf\"alische Wilhelms-Universit\"at M\"unster, Wilhelm Klemm Str.\ 9, 48149 M\"unster, Germany}
\affiliation{Center of Nonlinear Science (CeNoS), Westf{\"a}lische Wilhelms-Universit\"at M\"unster, Corrensstr.\ 2, 48149 M\"unster, Germany}
\affiliation{Center for Multiscale Theory and Computation (CMTC), Westf{\"a}lische Wilhelms-Universit\"at, Corrensstr.\ 40, 48149 M\"unster, Germany}
\email{u.thiele@uni-muenster.de}

\begin{abstract}
Employing a long-wave mesoscopic hydrodynamic model for the film height evolution we study ensembles of pinned and sliding drops of a volatile liquid that continuously condense onto a chemically heterogeneous inclined substrate. Our analysis combines on the one hand path continuation techniques to determine bifurcation diagrams for the depinning of single drops of nonvolatile liquid on single hydrophilic spots on a partially wettable substrate and on the other hand time simulations of growth and depinning of individual condensing drops as well as of the long-time behaviour of large ensembles of such drops. Pinned drops grow on the hydrophilic spots, depin and slide along the substrate while merging with other pinned drops and smaller drops that slide more slowly, and possibly undergo a pearling instability. As a result, the collective behaviour converges to a stationary state where condensation and outflow balance. The main features of the emerging drop size distribution can then be related to single-drop bifurcation diagrams.\\[3ex]
\textbf{\textcolor{red}{Published as:}} EPL 127, 54002 (2019). doi: 10.1209/0295-5075/127/54002
\end{abstract}
%
%\begin{keyword} 
% Ensemble of sliding drops \sep Heterogeneous substrates \sep
% Depinning and pearling \sep Condensation
%\end{keyword} 

\maketitle

\section{Introduction}
The behaviour of liquid drops on solid homogeneous and heterogeneous substrates is of high relevance to many processes of everyday life and for technological processes such as printing, coating and cooling \cite{GennesBrochard-WyartQuere2004}. The behaviour of individual drops is frequently studied experimentally and theoretically, considering, e.g., spreading and sitting drops without lateral driving \cite{Genn1985rmp}, laterally driven drops, e.g., by gravity on an incline, that are pinned by substrate heterogeneities \cite{BKHT2011pre,VFFP2013prl} or freely slide along a homogeneous substrate \cite{PoFL2001prl,EWGT2016prf}. %,Gree1978jfm,ChWh1992s,Pesc2018pf
However, in applications such as condensation or printing, one is often interested in the collective behaviour of large drop ensembles. This problem has attracted much interest in particular for rigid substrates where the interactions between individual drops and the resulting mass transfer processes determine the ensemble behaviour. The long-time merging within such drop ensembles is a particular soft matter example of a coarsening process similar to the Ostwald-ripening of crystalline nanoparticles \cite{RNRH2014n}, quantum dots \cite{VeGY2001s} or emulsion droplets \cite{Tayl1998acis} where the mean drop/cluster/dot size and their mean distance continuously increase in time following power laws.
For simple nonvolatile liquids on horizontal homogeneous substrates coarsening is well studied experimentally \cite{ABNP2002jfm,BLHV2012prl,BDHE2018prl} %,ZhBe1995l, RDWF1994jem,LiGr2003l, LiGr2002pre
and theoretically through simulations and asymptotic considerations \cite{GrWi2009pd,GORS2009ejam,KiWa2010jem} mainly based on thin-film (or lubrication or long-wave) equations with a mass-conserving dynamics \cite{OrDB1997rmp,CrMa2009rmp,Thie2010jpcm}. %GlWi2003pre,PiPo2004pf, OtRS2006sjma ,Kita2014ejam
Additionally including condensation, the process is also studied employing particle-based statistical models and Smoluchowski-type (cf.~\cite{Smol1916zp}) evolution equations for distribution functions of drop sizes \cite{Meak1992rpp}.
With lateral driving forces, the dynamics of drop ensembles is dramatically different as the sliding speed of drops strongly depends on their size. The resulting relative motion of differently sized drops makes overall coarsening much faster than without lateral driving forces. However, instabilities may counteract coalescence and at large times the ensemble dynamics may self-organise and converge to an almost stationary drop size distribution \cite{WTEG2017prl}. Examples are drops that slide under an air flow or on an incline as well as spinodal decomposition under flow \cite{HeNa1997ces}. Here, we investigate the influence of substrate heterogeneities and continuous condensation on the dynamics of laterally driven drop ensembles. We establish the resulting basic features employing a long-wave model and the particular choice of randomly distributed identical heterogeneities. Note that condensing and coalescing drops with instantaneous sliding avalanches have also been described with particle-based statistical models and Smoluchowski-type equations \cite{CRMF1989pra}.

More in detail, Ref.~\cite{EWGT2016prf} analyses a long-wave mesoscopic hydrodynamic model employing numerical path continuation techniques \cite{DWCD2014ccp,EGUW2019springer} and establishes the bifurcation behaviour of single sliding drops of nonvolatile liquid on smooth homogeneous inclined substrates. They find that at fixed lateral forcing [volume] beyond a critical volume [forcing] related to a saddle-node bifurcation and a nearby global bifurcation, sliding drops undergo a pearling instability \cite{PoFL2001prl} and emit satellite droplets at their back. Ref.~\cite{EWGT2016prf} also quantifies how sliding speed and the mentioned critical parameter values depend on drop size and driving strength. This allows one to characterise the fast coalescence of drops and the resulting fast coarsening under driving. 
In a multiscale approach, Ref.~\cite{WTEG2017prl} then connects the single-drop results with the time evolution of the drop size distribution obtained in large-scale direct numerical simulations (DNS) of drop ensembles and, in consequence, derives a Smoluchowski-type statistical model for the drop size distribution. Main features of the resulting steady distribution can be related to the bifurcation diagram 
%and stability properties 
of single sliding drops. The approach of Refs.~\cite{EWGT2016prf,WTEG2017prl} is based on a number of strong assumptions that are difficult to realise in experiments as most real substrates are heterogeneous, the used liquids are often volatile and periodic boundary conditions are rather difficult to achieve for sliding drops under lateral driving. Here, we adapt their approach to more realistic experimental conditions.

In particular, first, we incorporate (i) the deposition of liquid by condensation and (ii) heterogeneous wettability in the form of hydrophilic spots into the long-wave thin-film model. Next, we follow the methodology outlined above: We employ continuation techniques to obtain the bifurcation diagram for the depinning behaviour of single drops of nonvolatile liquid. This is then compared with simulations of growth and depinning dynamics of single drops that condense onto single spots. Finally, the resulting bifurcation diagram is related to large-scale DNS and it is discussed how the existence of heterogeneities and liquid condensation affect the ensemble behaviour.

\section{Modelling and numerical implementation}
We employ a nondimensional long-wave equation to model the time evolution of the height profile $h(x,y,t)$ that describes drops of a volatile liquid on a partially wetting, heterogeneous substrate,  cf.~\cite{ThKn2006njp,ToTP2012jem} and references therein:
\begin{equation}
\partial_t h = -\nabla\cdot\left[Q(h)\left(\nabla p + \boldsymbol{\chi}\right)\right] + \beta\left(p - \mu \right) 
\label{eq:TFEmodel} 
\end{equation}
with the pressure $p(x,y,t) = \Delta h + [1+ \xi g(x,y)]\Pi(h)$ where $\Delta h$ and $\Pi(h) = -\partial_h f(h)$ are the Laplace and Derjaguin (or disjoining) pressure, respectively \cite{Genn1985rmp,StVe2009jpm}. The latter results from the wetting energy $f(h) = -1/2h^2 + 1/5h^5$. Note that $p$ may be expressed as variation of a free energy functional \cite{Thie2018csa}. The function $\xi g(x,y)$ represents the heterogeneous wettability of the substrate, namely, the local long-wave equilibrium contact angle $\theta_\mathrm{eq}(x,y)\propto\sqrt{1+ \xi g(x,y)}$ while the scaled equilibrium adsorption layer height remains constant $h_0=1$. For drops on an incline, the driving force is given by $\boldsymbol{\chi}=G (\alpha,0)^T $ where $G$ is the gravitation number and $\alpha$ is the scaled inclination angle.%
\footnote{Starting from the dimensional Derjaguin pressure $\tilde \Pi(\tilde h) = - A/\tilde h^3 + B/\tilde h^6$, we scale height by $h_\mathrm{eq} = (B/A)^{1/3}$, lateral lengths by $l_0 = \sqrt{3}h_\mathrm{eq}/\sqrt{5}\theta_\mathrm{eq}$, and time by $t_0 = 9 \eta h_\mathrm{eq}/25\gamma \theta_\mathrm{eq}^4$. Then, $\theta_\mathrm{eq} = \sqrt{3 A/5 \gamma h_\mathrm{eq}^2}$ is the equilibrium contact angle at $\xi=0$ and $G = 3\rho g h^2_\mathrm{eq}/5 \gamma \theta_\mathrm{eq}^2$ is the gravitation number. The physical inclination angle is $\theta_\mathrm{eq}\alpha$. Here, we use $G = 10^{-3}$, e.g., for $\theta_\mathrm{eq}=0.1$, $h_\mathrm{eq}$ is about a micron. We do not expect qualitative changes for smaller $h_\mathrm{eq}$ \cite{EWGT2016prf} - a case that is at present not numerically feasible.} Here, the heterogeneities take the form of randomly distributed identical small circular hydrophilic regions, i.e., more wettable spots, with a small continuous transition region towards the partially wetting background substrate. In particular, for a single spot we employ $g(\tilde r)= -\frac{1}{2}[\tanh(\tilde r + R) - \tanh(\tilde r - R)]$ with $\tilde r^2= (x-x_i)^2 + (y-y_i)^2$, where $R$ is the uniform spot radius and $(x_i,y_i)$ the centre position of spot $i$.  Furthermore, $\beta$ is an evaporation rate and $\mu$ is the partial ambient vapour pressure. In combination they control the strength of condensation or evaporation. Note that the dependence on pressure automatically incorporates the Kelvin effect and a wettability-dependence of phase change -- for a discussion of evaporation models see \cite{Thie2014acis}.  The model is analysed employing (i) numerical pseudo-arclength path-continuation techniques \cite{DWCD2014ccp,EGUW2019springer} implemented using pde2path \cite{UeWR2014nmma} and (ii) direct numerical simulations (DNS) based on a finite-element method on a quadratic mesh with bilinear ansatz functions and a 2nd-order implicit Runge-Kutta scheme for time stepping implemented using the DUNE PDELab framework \cite{BBDE2008c,BBDE2008cb}.

\begin{figure} \center
\includegraphics[width=0.7\hsize]{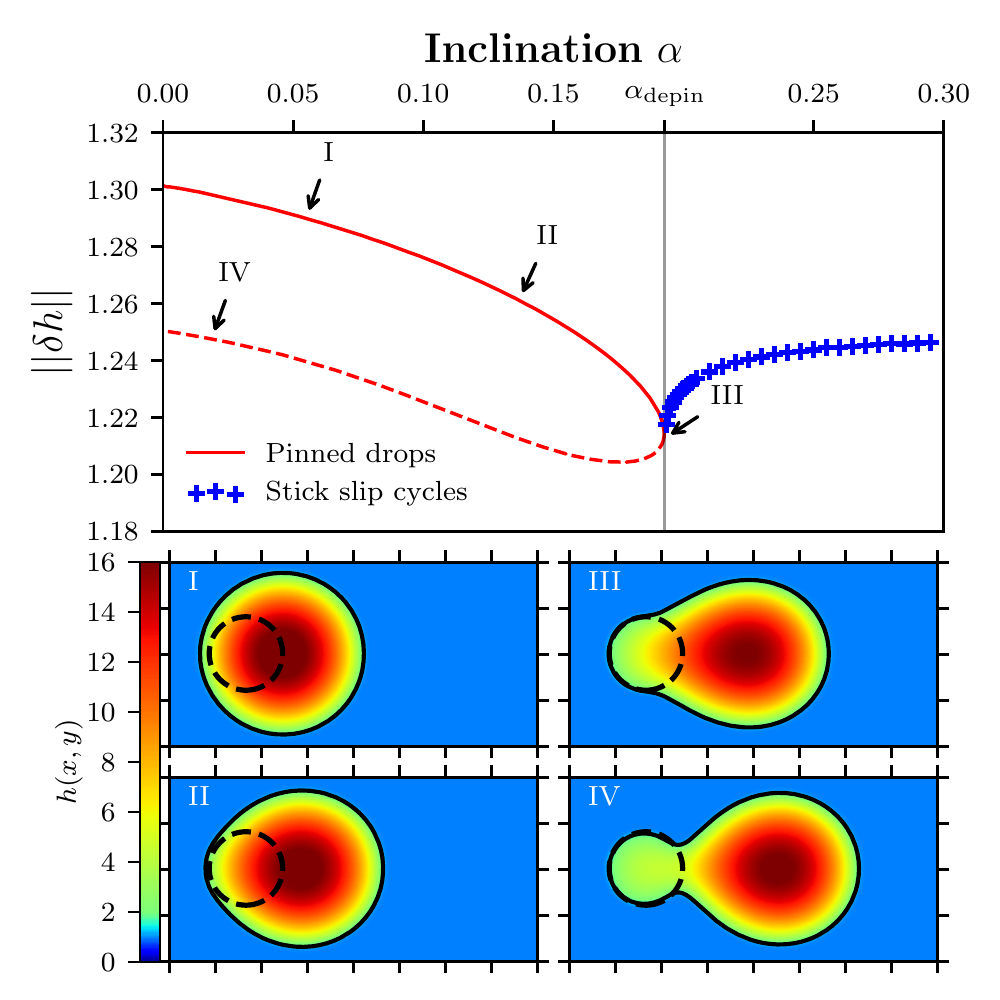}
\caption{Typical bifurcation diagram (top) and selected drop profiles as contour plots (bottom) related to the depinning of a drop of partially wetting, nonvolatile ($\beta=0$) liquid from a circular hydrophilic spot (dashed lines in bottom panels) under lateral driving. The bifurcation diagram gives the (time-averaged) $L^2$-norm $||\delta h||$ as a function of substrate inclination $\alpha$ (i.e., strength of driving) for linearly stable and unstable pinned drops (solid and dashed line, respectively) and for the depinned sliding drops that undergo a periodic stick-slip motion in the considered periodic setting (cross symbols). The parameters of the drop profiles are indicated by corresponding roman numbers in the upper panel. The domain size is $l_x\times l_y = 200\times100$ and $\xi = 1.0$, the drop volume is fixed at $V_\mathrm{D} = 5\times10^4$ and the spot radius is $R = 20$.}
\label{fig:dynhe_alpcont_2D}
\end{figure}

\section{Single-drop depinning}
On a smooth homogeneous substrate, drops of any size slide for arbitrarily small lateral driving, i.e., for any $\alpha\neq0$ \cite{EWGT2016prf}. In stark contrast, on a heterogeneous substrate, drops are pinned at small driving strength as investigated in depth with long-wave models for drops on one-dimensional substrates \cite{ThKn2006prl,ThKn2006njp} and on two-dimensional substrates with stripe-like heterogeneity \cite{BeHT2009el,BKHT2011pre}. Fig.~\ref{fig:dynhe_alpcont_2D} presents for a single drop of nonvolatile liquid pinned by a single circular hydrophilic spot a typical bifurcation diagram at fixed drop volume $V_\mathrm{D}$ employing the driving strength $\alpha$ as control parameter (top)\footnote{%
As solution measure we mainly use the $L^2$-norm
$||\delta h|| := \sqrt{\Omega^{-1}\int_{\Omega} \left[h/h_0 -
    1\right]^2 \,dxdy}$. Spherical cap-like drops of large
volume are characterised by a relatively large $||\delta h||$, which
is reduced for drops that are small or strongly deformed. The drop volume $V_\mathrm{D}$ is measured as the volume above the adsorption layer of height $h_0=1$.} and selected corresponding drop profiles (bottom). Further cases are discussed in % Sec.~\ref{sec-app-pin} 
the Supplementary Material (sections~\ref{sec:supp-pin} to~\ref{sec:supp-elongated}).
At small driving there exists a branch of linearly stable pinned drops sitting off-centre on the spot (e.g., I, II) and a branch of unstable drops that are located slightly downstream of the spot and connect to it by a narrow liquid bridge (e.g., IV). Starting from a spherical cap-like drop at $\alpha=0$ (not shown), with increasing $\alpha$ the stable drop first keeps its spherical cap-like shape but shifts its centre downstream (I). Further increasing $\alpha$, the drop is increasingly deformed, so that $||\delta h||$ decreases monotonically (II). The branch of linearly stable states ends in a saddle-node bifurcation at $\alpha_\mathrm{depin} \approx 0.1926$ (III) where it annihilates with the branch of unstable states. As known from other geometries \cite{ThKn2006prl,BKHT2011pre}, at the saddle-node bifurcation a branch of stick-slip states emerges in a global bifurcation. As here we work with periodic boundary conditions, these represent time-periodic states with a period that diverges when approaching the bifurcation point.

Each cycle of the resulting motion has two distinct phases: first, the drop is pinned by the spot but slowly stretches downstream. Then it depins and slides fast to the next defect where it pins again. This is illustrated in the Supplementary Material  (section~\ref{sec:supp-slip}). Close to the bifurcation, the time scales for the stick- and  the slide-phase strongly differ, and the overall behaviour closely resembles experimentally observed stick-slip motion \cite{VFFP2013prl}. Note, that the unstable steady states represent critical perturbations that have to be overcome to depin and start to slide already below the critical driving strength, i.e., for $\alpha<\alpha_\mathrm{depin}$.

\section{Single drop condensation and depinning}
\begin{figure} \center
\includegraphics[width=0.8\hsize]{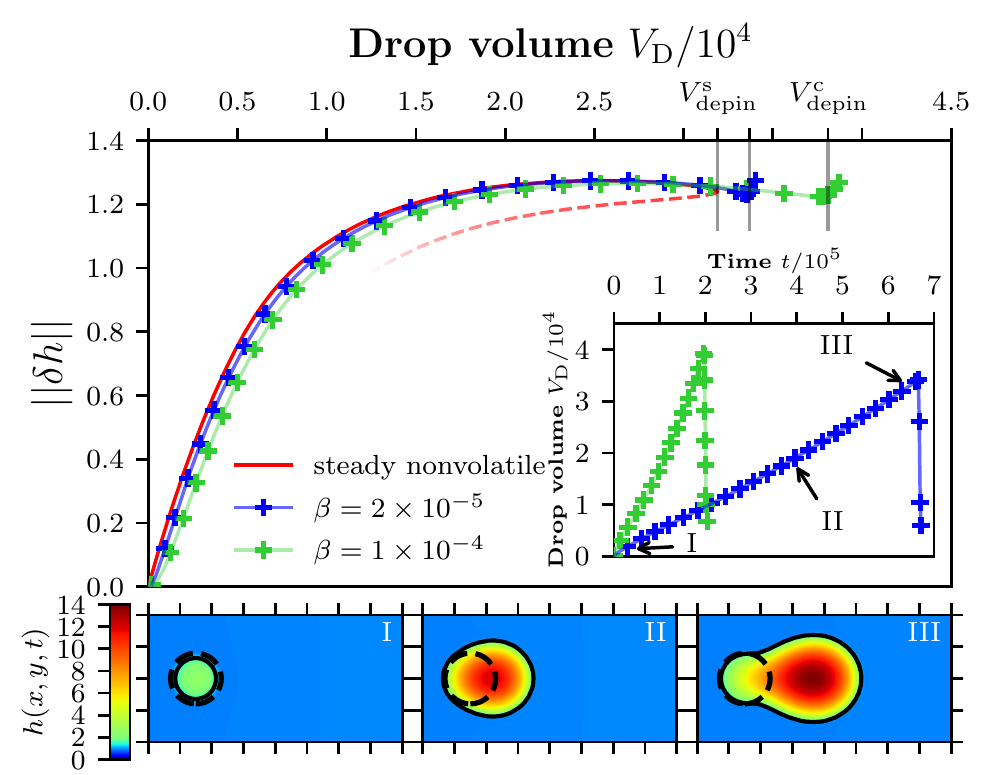}
\caption{(top) The lines with symbols characterise the time evolution of a single drop of volatile liquid that grows by condensation on a hydrophilic spot. Shown is the $L^2$-norm $||\delta h||$ over drop volume $V_\mathrm{D}$ for two different condensation rates as given in the legend and for comparison the bifurcation curve of pinned nonvolatile drops (bare solid line). The short vertical lines indicate the volume at depinning in the nonvolatile ($V^\mathrm{s}_\mathrm{depin}$, left line) and volatile ($V^\mathrm{c}_\mathrm{depin}$, right lines) case. The inset gives $V_\mathrm{D}(t)$ for the condensing drops. The bottom row gives snapshots of the growing pinned drop for $\beta = 2\times10^{-5}$ at times marked by roman numbers in the inset.
The domain size is $l_x\times l_y = 200\times100$, $\xi = 1.0$, the inclination is fixed at $\alpha = 0.3$, the spot radius is $R = 20$ and the partial vapour pressure that drives condensation is $\mu = -0.05$.}
\label{fig:dynhe_depin_single_cond}
\end{figure}

Next we introduce condensation ($\beta>0$ and $\mu<0$ in Eq.~(\ref{eq:TFEmodel})). Then, on the partially wettable background substrate the equilibrium adsorption layer height is only slightly shifted (given by $\Pi(h)-\mu=0$), but on the hydrophilic spots, the film height grows with a rate $|\beta \mu|$ (marginally slowed down by the Kelvin effect). As a result, individual drops condense onto the hydrophilic substrate defects. As their mass continuously grows, they eventually reach the critical mass for depinning at fixed inclination and depin under the influence of the lateral driving force.  After depinning, drops slide and may undergo a pearling instability similar to Ref.~\cite{WTEG2017prl}. 

We first quantify this process and its dependence on condensation rate in Fig.~\ref{fig:dynhe_depin_single_cond} for a single drop on a hydrophilic spot. The figure compares the bifurcation curve of steady pinned drops as a function of their volume $V_\mathrm{D}$ (at fixed inclination) with the time evolution of condensing drops for two different condensation rates. Note, that in contrast to Fig.~\ref{fig:dynhe_alpcont_2D} periodic boundary conditions (BC) are only used in the spanwise (i.e., $y-$) direction, while in streamwise (i.e., $x-$) direction Neumann BC are used. At the upstream border the film is always flat and Neumann BC result in no-flux BC while downstream these BC allow drops to slide out of the domain. The same BC are employed in the ensemble DNS below.

Here, each time simulation is started from a flat film of adsorption layer height $h_0= 1$, i.e., $V_\mathrm{D} = 0$ and $||\delta h||=0$. Subsequently, liquid condenses into a drop on the ideally wettable spot. As soon as the height profile deviates from a flat film, a finite Laplace pressure results in a slight decrease [increase] of condensation in the bulk drop [contact line] region. This results in further small internal fluxes that rearrange liquid within the drops. %As the drop grows, this additional influence decreases at the drop centre.

Inspecting the top panel of Fig.~\ref{fig:dynhe_depin_single_cond} in detail, one appreciates that the growing drops (e.g., bottom panels II and III) closely follow in the $(V_\mathrm{D}, ||\delta h||)$-plane the bifurcation curve representing stable steady drops of different volumes. This holds up to the saddle-node bifurcation that indicates depinning for drops of nonvolatile liquids. The slightly smaller norm at identical volume indicates a smaller contact angle -- an effect that is more pronounced at larger condensation rates, i.e., at larger deviation from equilibrium. For evaporating drops, it is known that due to evaporation-driven internal flows towards the contact line region, the contact angle is larger than the equilibrium value \cite{Morr2001jfm,ToTP2012jem}. Here, we encounter the expected opposite effect for condensing drops due to condensation-driven internal flows towards the drop centre.

When the drops pass the critical volume for depinning $V_\mathrm{depin}^\mathrm{s}\approx 3.19\times10^4$ of the steady nonvolatile drop, they depin. However, with ongoing condensation, the volume where this happens is moderately shifted to a larger $V_\mathrm{depin}^\mathrm{c}$ (in Fig.~\ref{fig:dynhe_alpcont_2D} indicated by the two short vertical lines on the right) because the time scale of the depinning process has to become shorter than the one for condensation. Therefore the shift $V_\mathrm{depin}^\mathrm{c}-V_\mathrm{depin}^\mathrm{s}$ is larger for faster condensation (larger $\beta$). After the connection to the defect is capped, at the present moderate lateral driving the sliding drop closely approaches a slightly oval spherical cap-like shape (small but distinct increase of the norm close to $V_\mathrm{depin}^\mathrm{c}$).
The sliding drop moves downstream and quickly leaves the domain. This results in the abrupt decrease in volume visible in the inset of Fig.~\ref{fig:dynhe_depin_single_cond}. As the qualitative behaviour is similar for all considered condensation rates, now we focus on the larger one ($\beta=10^{-4}$) as this allows for large-scale DNS on time scales that are large as compared to time scales of condensation, depinning and sliding.

\begin{figure} \center
\includegraphics[width=0.7\hsize]{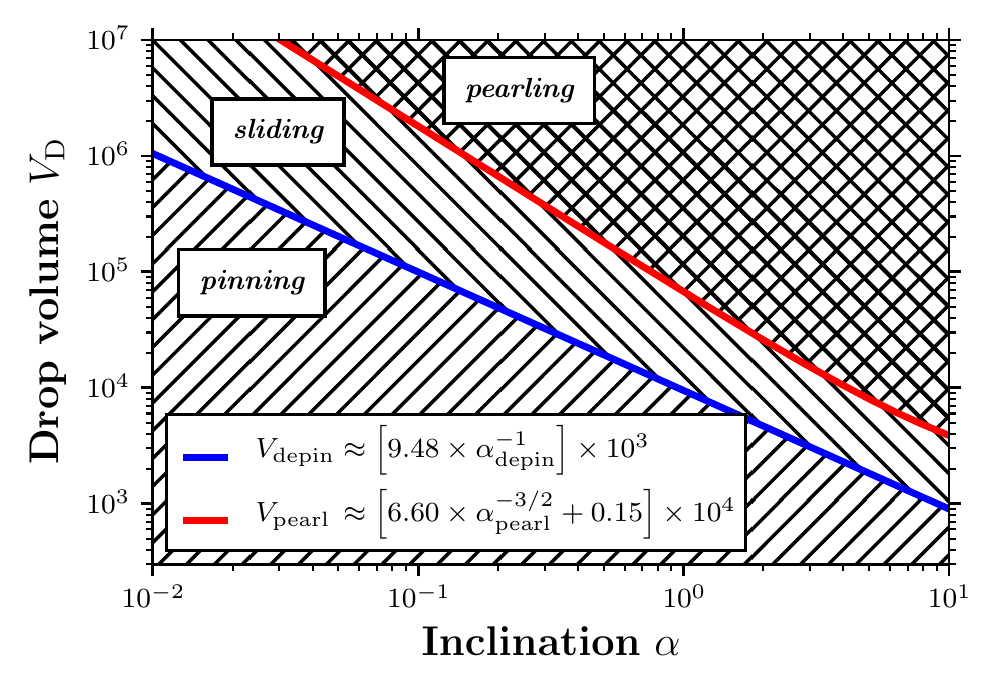}
\caption{Morphological phase diagram in the nonvolatile case indicating where pinned, stable sliding and pearling drops dominate in the parameter plane spanned by drop volume and inclination angle. The borders between regions correspond to power laws (given in the inset) extracted from sets of bifurcation diagrams as, e.g., Fig.~\ref{fig:dynhe_alpcont_2D} above and Fig.~1 of \cite{EWGT2016prf}.
Also see Supplementary Material (section~\ref{sec:supp-power}). Remaining parameters are as in Fig.~\ref{fig:dynhe_alpcont_2D}. }
\label{fig:conc_phase_space}
\end{figure}

If the domain is sufficiently extended, at high driving strength one can observe that the drop undergoes a pearling instability (as in Ref.~\cite{EWGT2016prf}). As in the following, depinning and pearling will play an important role, we present in Fig.~\ref{fig:conc_phase_space} a morphological phase diagram for single drops in the nonvolatile case. It indicates in a log-log plot volume and inclination ranges where drops are pinned at the defect, slide down the homogeneous background substrate and undergo a pearling instability while sliding. The separating lines can be fitted by the power laws given in the legend of Fig.~\ref{fig:conc_phase_space}. Also see Supplementary Material (section~\ref{sec:supp-power}).

\begin{figure*} \center
\includegraphics[width=0.8\hsize]{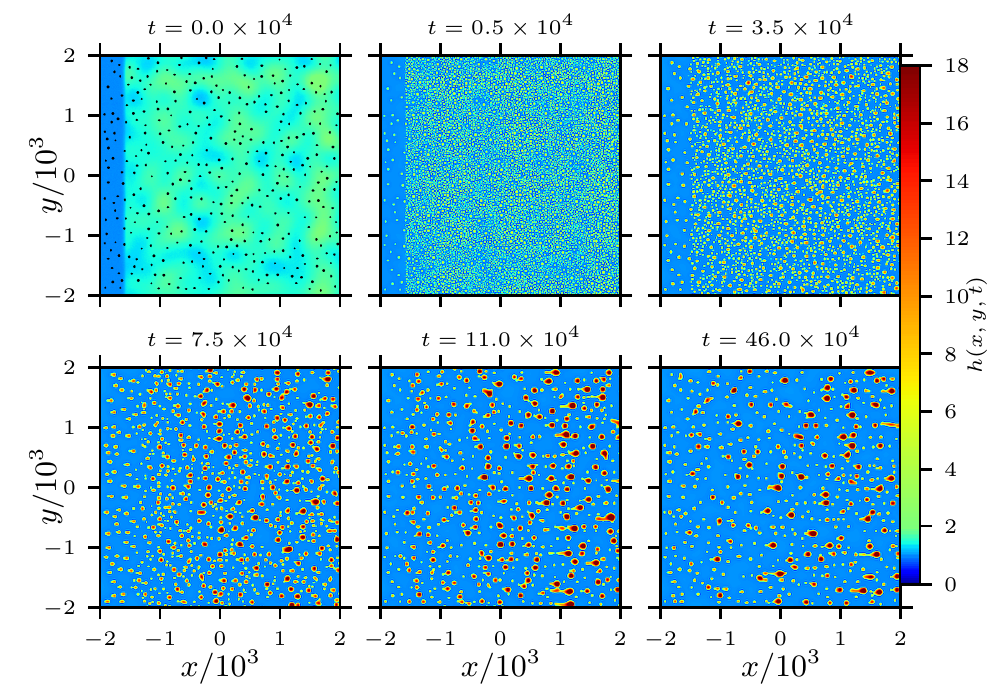}
\caption{Shown are snapshots from a large-scale direct time simulation [Eq.~(\ref{eq:TFEmodel})] of an ensemble of condensing drops on an inclined substrate with $N_\mathrm{S}\approx400$ identical randomly distributed hydrophilic spots with $R=20$ (black dots in top left panel). 
The condensation rate is moderate $\beta = 10^{-4}$, $\alpha = 0.5$, and the domain size is $4000\times 4000$. During an initial transient, spinodal dewetting contributes to the formation of drops that later (from about $t = 10^4$) mainly condense onto the hydrophilic spots. From $t \approx 7.5\times10^4$ the dynamics is dominated by pinned and sliding drops. In the long-time limit (here, reached at $t \approx 25\times10^4$) the dynamics converges to a stationary state where condensation, depinning and inclination-driven outflow balance, resulting in a steady drop size distribution.
}
\label{fig:dynhe_ense_tseries}
\end{figure*}

\section{Large-scale time simulation}
Large-scale DNS of Eq.~(\ref{eq:TFEmodel}) are conducted on a large spatial domain ($4000\times4000$) with about 400 identical randomly distributed, not overlapping, hydrophilic spots of radius $R=20$ (see black spots in the top left panel of Fig.~\ref{fig:dynhe_ense_tseries}) for different fixed inclination angles. We believe this relatively high spot density presents a good first approximation for a heterogeneous substrate. Note that at this density most drops will interact with several defects before leaving the domain. This also ensures that the results will not critically depend on streamwise domain size as may be the case at low spot density.

Statistical analyses are applied to the resulting ensembles of growing pinned and sliding drops.\footnote{To quantify the process, the total number of drops $N_\mathrm{D}(t)$ in the domain is determined as well as all individual drop volumes and the resulting drop size distribution $f(V_\mathrm{D},t)$. We define an individual drop via the connected area $A_\mathrm{D}$ of its footprint where the height $h(x,y,t)$ is larger than a threshold height that is slightly larger than the height of the adsorption layer (here $h_\mathrm{thresh} = 1.05$).
For each step of the DNS, all drop volumes $V_\mathrm{D}$ are calculated by integrating $h(x,y,t)$ over the corresponding $A_\mathrm{D}$. Then the distribution $f(V_\mathrm{D},t)$ is obtained by a Gaussian kernel density estimate (KDE) \cite{Scott2015}.} 
To shorten the initial transient, here, the initial condition is a flat film of height $h_\mathrm{ini} = 2.0$ perturbed by small-amplitude additive noise and a further spatial harmonic modulation of large wavelength. 
The latter induces different initial conditions (IC) at the individual hydrophilic spots, so that artificially synchronised behaviour is avoided and the system sufficiently fast approaches a purely statistical state. Furthermore, at the upstream boundary a strip of bare adsorption layer height is introduced into the IC to ensure the no-flux BC (Fig.~\ref{fig:dynhe_ense_tseries}(top left)). In this way, the total volume in the domain exclusively results from a balance of condensation and downstream outflow. 

\begin{figure} \center
\includegraphics[width=0.8\hsize]{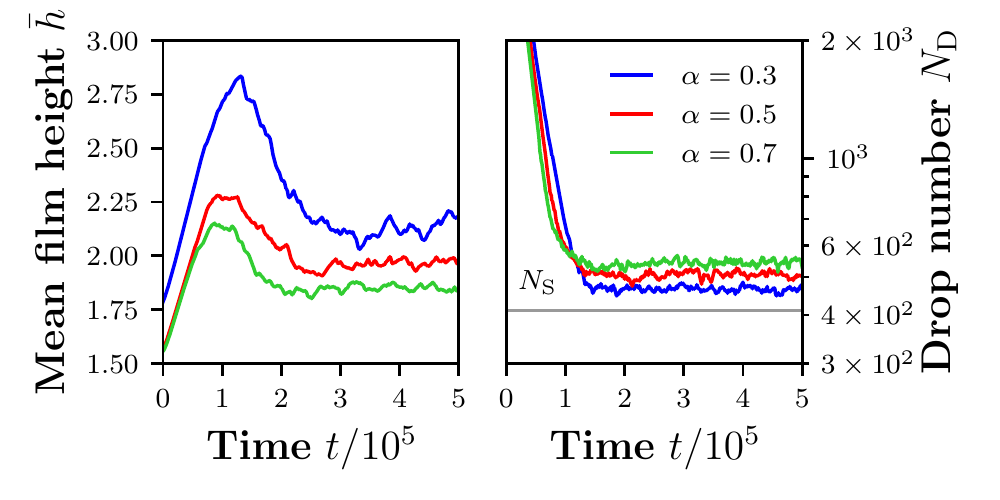}
\caption{Time evolution of (left) the mean film height $\bar{h}$ and (right) the drop number $N_\mathrm{D}$ obtained in large-scale DNS at different inclinations $\alpha$ as given in the legend. The number $N_\mathrm{S}$ of hydrophilic spots is shown as thin horizontal line. In all cases, first ($t \lesssim 10^5$) condensation, spinodal dewetting and drop coarsening dominate, i.e.,  $\bar{h}$ increases and $N_\mathrm{D}$ decreases. Then depinned drops slide out of the domain and $\bar{h}$ decreases, until at about $t = 2\dots3\times10^5$ a balance of condensation and outflow is established. With decreasing $\alpha$, the stationary state is characterised by a smaller $N_\mathrm{D}$ and a larger $\bar{h}$.}
\label{fig:dynhe_ense_h0_noofdrops}
\end{figure}

The series of snapshots in Fig.~\ref{fig:dynhe_ense_tseries} presents important phases of the resulting dynamics for $\alpha=0.5$. A comparison with other inclinations is given in the Supplementary Material (section~\ref{sec:supp-ensemble}). The corresponding dependencies of mean film height $\bar{h}$ and drop number $N_\mathrm{D}$ on time are given in Fig.~\ref{fig:dynhe_ense_h0_noofdrops}. The first phase (top row of Fig.~\ref{fig:dynhe_ense_tseries}) represents a transient dominated by spinodal dewetting that results in the fast emergence of many small droplets and their subsequent coarsening (decrease of $N_\mathrm{D}$ in Fig.~\ref{fig:dynhe_ense_h0_noofdrops}) accompanied by an ongoing increase of $\bar{h}$ due to condensation. 
The effect of the hydrophilic spots is clearly visible at $t = 0.5\times10^4$ (Fig.~\ref{fig:dynhe_ense_tseries}) where significantly larger droplets have developed on all of them. They absorb the smaller droplets within their immediate vicinity and attract most condensation. The remaining small droplets continue their coarsening and fusion into the large drops at the defects. A clear qualitative difference is seen in the transition from $t = 3.5\times10^4$ to $t = 7.5\times10^4$ as most droplets from initial dewetting have disappeared and the dynamics is dominated by condensation and depinning. At $t \approx 10^5$ the decreasing $N_\mathrm{D}$ is converging to a steady number while $\bar{h}$ still decreases due to the outflow of the initial batch of larger drops.

\begin{figure} \center
\includegraphics[width=0.7\hsize]{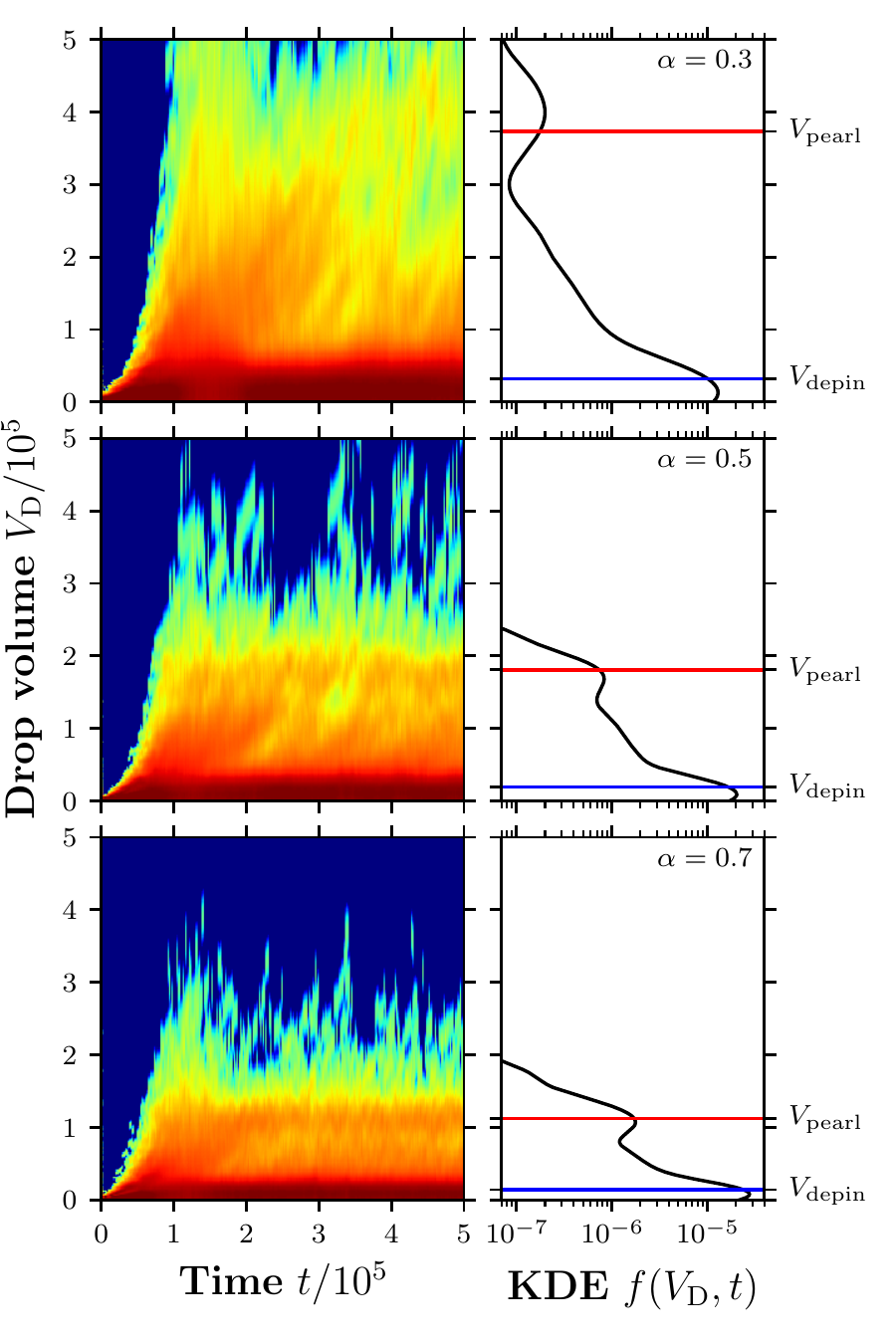}
\caption{The left panels show time evolutions of the drop size distribution as space-time plots of the Gaussian kernel density estimate (KDE) $f(V_\mathrm{D},t)$ for (top) $\alpha = 0.3$, (middle) $\alpha = 0.5$ and (bottom) $\alpha = 0.7$. The phases described at Fig.~\ref{fig:dynhe_ense_h0_noofdrops} can be well distinguished. The respective right panels give the final steady drop size distributions $f(V_\mathrm{D})$, obtained as time average of the converged but fluctuating distribution from $t=4.5\times 10^5$ to $t=5.0\times 10^5$. The critical drop sizes for depinning $V_\mathrm{depin}$ and pearling $V_\mathrm{pearl}$ are indicated as horizontal lines (cf.~Fig.~\ref{fig:conc_phase_space}).
}
\label{fig:dynhe_ense_KDEtime}
\end{figure}
 
At large times (e.g., $t = 46.0\times10^4$ in Fig.~\ref{fig:dynhe_ense_tseries}) a steady drop size distribution has developed where $N_\mathrm{D}$ and $\bar{h}$ fluctuate about their respective mean values. This implies that condensation and outflux balance in a stationary state. Thereby, the values of the corresponding plateaus in Fig.~\ref{fig:dynhe_ense_h0_noofdrops} for $N_\mathrm{D}$ [$\bar{h}$] decrease [increase] with increasing inclination: At low $\alpha$, the depinning threshold $V_\mathrm{depin}$ is larger than at high $\alpha$, and the ensemble consists of fewer and larger drops (for an image see Supplementary Material, section~\ref{sec:supp-ensemble}). In all cases, the number of hydrophilic spots $N_\mathrm{S}$ naturally forms the lower limit for $N_\mathrm{D}$, such that $N_\mathrm{D}-N_\mathrm{S}$ indicates the number of sliding drops. Due to later depinning and slower sliding the mean height in the domain is larger at lower $\alpha$. It is notable that here $N_\mathrm{D}$ truly converges in the long time limit while in the nonvolatile case on homogeneous substrates the drop number still slowly decreases in the long-time limit \cite{WTEG2017prl}. There this small drift is due to large linearly stable sliding drops that feature a long backwards protrusion \cite{EWGT2016prf}. In the present case, such drops are disturbed and broken up by the heterogeneities.

Two further effects are visible in Fig.~\ref{fig:dynhe_ense_tseries}: First, one discerns a gradient in drop sizes in streamwise direction which results from the increase of drop size as they move through the domain and collect liquid from hydrophilic spots that they pass. Second, in contrast to the homogeneous substrates \cite{WTEG2017prl}, the ensemble always remains dominated by a large number of relatively small drops pinned at the hydrophilic spots. This is very clear in Fig.~\ref{fig:dynhe_ense_KDEtime} where on the left volume-time plots of the drop size distribution $f(V_\mathrm{D},t)$ are shown for different driving strength while on the right the resulting steady distributions $f(V_\mathrm{D})$ are shown. We always find that a characteristic double-peaked drop size distribution emerges. The loci of the two peaks are close to the critical drop sizes for depinning $V_\mathrm{depin}$ and pearling $V_\mathrm{pearl}$, respectively, that are obtained from the single-drop bifurcation diagrams (Figs.~\ref{fig:dynhe_alpcont_2D} and~\ref{fig:conc_phase_space}). With decreasing $\alpha$ the peaks become wider and their distance becomes larger. Notably, at low $\alpha$ an intermediate range between the peaks emerges where $f(V_\mathrm{D})$ decreases exponentially with increasing drop size.

\section{Conclusion} We have employed a long-wave film height evolution equation to qualitatively study the collective behaviour of ensembles of pinned and sliding drops of volatile liquid on chemically heterogeneous inclined substrates combining path-continuation methods and large scale direct numerical simulations. We have obtained bifurcation diagrams that quantify the depinning of individual drops from hydrophilic spots in the nonvolatile case and could show that such a bifurcation curve roughly predicts the path taken by the continuous condensation of individual drops onto such spots. In the case of the drop ensemble condensing onto many identical randomly distributed, not overlapping, hydrophilic spots, beyond depinning the drops slide along the substrate, collect the liquid of other pinned drops and of smaller drops that slide more slowly. Sufficiently large sliding drops undergo a pearling instability. We have found, that as a result of these competing processes the collective behaviour of the drop ensemble converges to a stationary state of on average balanced condensation and outflow. The stationary state is characterized by a drop size distribution whose main features are related to the single-drop bifurcation diagrams.

In the future, it will be interesting to determine how the collective behaviour depends on details of the wetting behaviour, on the parameters related to condensation and on the characteristics of the substrate heterogeneities. Of particular interest are the spatial, size, shape and wettability distribution of the heterogeneities. Further it is of interest if the descibed features are robust beyond the long-wave approximation.

\section*{Acknowledgement} We acknowledge support by the German-Israeli Foundation for Scientific Research and Development (GIF, Grant No.\ I-1361-401.10/2016) and by the Deutsche Forschungsgemeinschaft (DFG, Grant No.\ TH781/12 within SPP~2171).

\section{Supplementary Material\protect\footnote{In contrast to the published version, here we include the material in the same document.}}
\label{sec:supp}
Here we provide further detailed information supplementing the main text on the collective behaviour of ensembles of condensing liquid drops on heterogeneous inclined substrates.  The analysis there combines studies of the depinning of single drops from hydrophilic spots on a partially wettable background and of the long-time behaviour of large drop ensembles. The Supplementary Material provides on the one hand further detailed bifurcation diagrams for the single-drop case. On the other hand, we briefly discuss the influence of the inclination angle on the ensemble behaviour at constant condensation rate.

\subsection{Model}
As in the main text, we employ a nondimensional long-wave equation to model the time evolution of the height profile $h(x,y,t)$ that describes drops of a volatile liquid on a partially wetting, heterogeneous substrate,  cf.~\cite{ThKn2006njp,ToTP2012jem}:
\begin{equation}
\partial_t h = -\nabla\cdot\left[Q(h)\left(\nabla p + \boldsymbol{\chi}\right)\right] + \beta\left(p - \mu \right) 
\label{eq:TFEmodel} 
\end{equation} 
with the pressure $p(x,y,t) = \Delta h + [1+ \xi g(x,y)]\Pi(h)$ where $\Delta h$ and $\Pi(h) = -\partial_h f(h)$ are the Laplace and Derjaguin (or disjoining) pressure, respectively \cite{Genn1985rmp,StVe2009jpm}. The latter results from the wetting energy $f(h) = -1/2h^2 + 1/5h^5$. Note that $p$ may be expressed as variation of a free energy functional \cite{Thie2018csa}. The function $\xi g(x,y)$ represents the heterogeneous wettability of the substrate, namely, the local long-wave equilibrium contact angle $\theta_\mathrm{eq}(x,y)\propto\sqrt{1+ \xi g(x,y)}$ while the scaled equilibrium adsorption layer height remains constant $h_0=1$. For drops on an incline, the driving force is given by $\boldsymbol{\chi}=G (\alpha,0)^T $ where $G$ is the gravitation number and $\alpha$ is the scaled inclination angle.

\begin{figure}
\centering
\includegraphics[width=0.55\hsize]{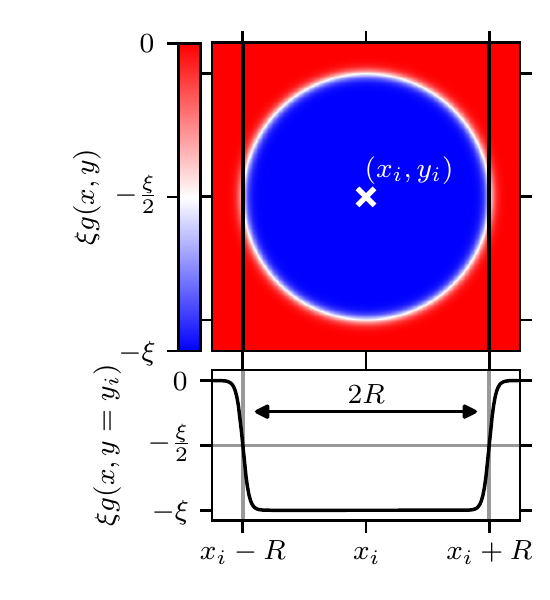}
\caption{(bottom) One-dimensional and (top) two-dimensional structure function of the wetting modulation. For two-dimensional drops a circular area with radius $R$ of higher wettability (blue) is embedded in an homogeneous background of partial wettability (red). The transition is continuous and the transition zone is of fixed width. The cross-section in the $x$-direction illustrates the continuous transition and also defines the structure function of the one-dimensional problem.}
\label{fig:sketch_spot_heterogeneity}
\end{figure}

The heterogeneities take the form of small circular hydrophilic regions, i.e., more wettable spots, with a small continuous transition region towards the partially wetting background substrate. In particular, for a single spot we employ $g_I= -\frac{1}{2}[\tanh(\tilde r + R) - \tanh(\tilde r - R)]$ with $\tilde r^2= (x-x_i)^2 + (y-y_i)^2$, $R$ the spot radius and $(x_i,y_i)$ the position of the centre of spot $i$. A sketch of an individual defect is given in Fig.~\ref{fig:sketch_spot_heterogeneity}.

\subsection{Depinning of single drop}
\label{sec:supp-pin}
\begin{figure}
\centering
\includegraphics[width=0.7\hsize]{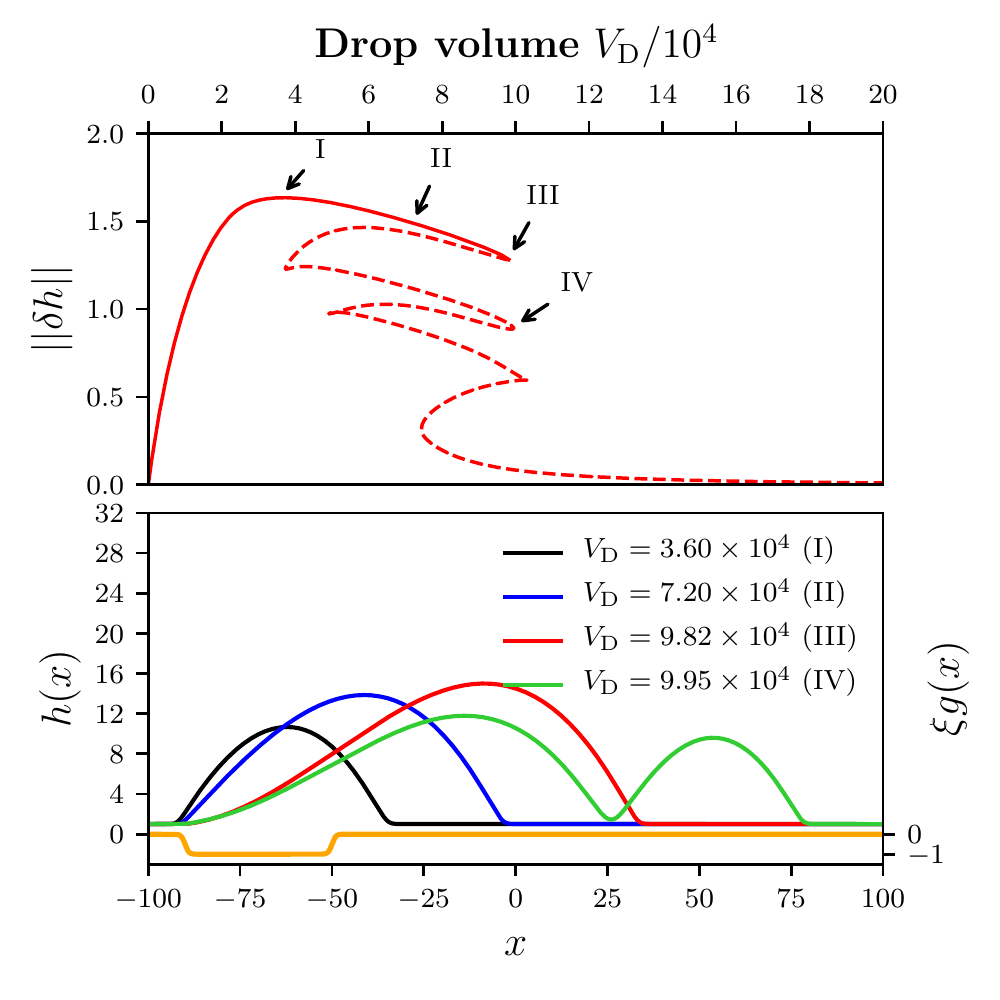}
\caption{(top) Bifurcation diagram and (bottom) height profiles of
  pinned one-dimensional drops on a heterogeneous substrate, i.e.,
  transversally invariant ridges on a two-dimensional substrate with a
  stripe-like heterogeneity. The heterogeneity in wettability has an
  extension of $2R$ with $R=20$ and is shown in the lower panel as orange line. The bifurcation diagram shows the $L^2$-norm $||\delta h||$ as a function of drop volume $V_\mathrm{D}$ at fixed inclination $\alpha = 0.3$. It indicates the existence of a branch of stable pinned drops (solid line) and of several branches of unstable steady drop states (dashed lines). The lower panel gives selected profiles as indicated by roman numbers.}
\label{fig:dynhe_volcont_1D}
\end{figure}

First, we consider drops in a one-dimensional (1d) system with a hydrophilic patch of width $2R = 40$. Note that the 1d drops correspond to transversally invariant ridges in a two-dimensional (2d) system that sit on hydrophilic stripes. Employing continuation techniques to study pinned drops at fixed inclination $\alpha = 0.3$ but varying drop volume $V_\mathrm{D}$ we obtain the bifurcation diagram in Fig.~\ref{fig:dynhe_volcont_1D}.  Starting at zero volume, i.e., at a flat film of height $h= 1$ we obtain the primary solution branch that starts with linearly stable drops (solid line) as illustrated by profiles I and II. With increasing volume, the drops covers not only the hydrophilic patch but increasingly larger areas of the downstream partially wetting substrate. For small ridge volumes $V_\mathrm{D} \lesssim 3.60\times 10^4$ the $L^2$-norm $||\delta h||$ increases monotonically and the drop profiles only slightly deviate from the symmetric parabolic form that they have without lateral driving force in the employed long-wave approximation (cf.~Refs.~\cite{TBBB2003epje,ThKn2006njp,HLHT2015l}). After reaching a maximum, $||\delta h||$ decreases with increasing volume and the height profile becomes increasingly asymmetric, i.e., the contact angles at front and rear become clearly different. 

At a characteristic saddle-node bifurcation at $V_\mathrm{D} \approx 9.82\times10^4$ which corresponds to a critical pinned drop (III), the curve folds back and becomes unstable. Several more saddle-node bifurcations follow, making the steady drop states more and more unstable. An example for such a drop is profile IV that features a main drop on the hydrophilic part and a smaller downstream satellite. A further steady satellite drop emerges at the next saddle-node (not shown).  After a final saddle-node at large volumes the steady state approaches a modulated film in analogy to results obtained in Ref.~\cite{ThKn2006njp} for another Derjaguin pressure and other heterogeneities.

\begin{figure} \center
\includegraphics[width=0.7\hsize]{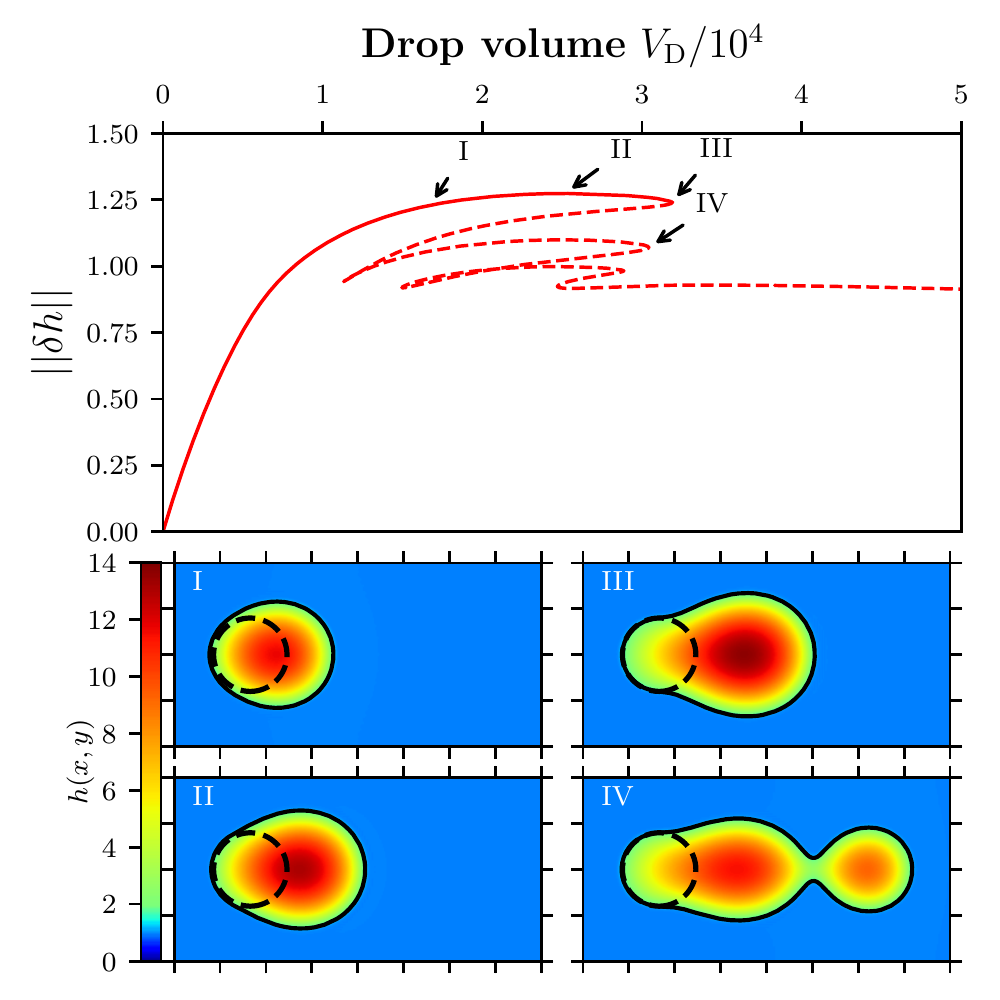}
\caption{Bifurcation diagram in the volume and height profiles of the pinned  two-dimensional drop solutions for a fixed substrate inclination $\alpha = 0.3$. The dotted circular line shows the geometry of the hydrophilic spot. Analogous to Fig.~\ref{fig:dynhe_volcont_1D} a stable solution branch of pinned drops is obtained from a flat film. Due to the driving force, the drops increasingly cover areas of the surrounding substrate of low wettability with increasing volume (I,II). The contact line at the back of the drop is deformed by the heterogeneity until at $V_\mathrm{D} \approx 3.19\times 10^4$ a saddle-node bifurcation occurs (III). Beyond the first saddle-node bifurcation, unstable solutions are found that show steady satellite droplets connected by thin bridges to the main drop (IV). The domain size is $l_x\times l_y = 200\times100$ and $\xi = 1.0$, and the spot radius is $R = 20$.}
\label{fig:dynhe_volcont_2D}
\end{figure}

Next, we consider the 2d case where the heterogeneity consists of a hydrophilic spot as in the top panel of Fig.~\ref{fig:sketch_spot_heterogeneity} and in all the main paper. The  top panel of Fig.~\ref{fig:dynhe_volcont_2D} gives the bifurcation diagram using drop volume $V_\mathrm{D}$ as control parameter and fixing inclination at $\alpha = 0.3$. Typical height profiles are presented in the lower panels.
The bifurcation curve has an overall shape similar to the one for the 1d drops. However, the solutions correspond to intrinsically 2d drops resulting from the spot-like geometry of the heterogeneity. Looking at the height profiles along the stable branch, the droplets with small volumes almost correspond to spherical caps (I), which only show a slight deformation in the back region due to the interaction of contact line and wetting heterogeneity (II).
The contact line adheres to the transition between the hydrophilic spot and the surrounding substrate with lower wettability, so that the shape of the heterogeneity is reflected in the deformation. Also for the 2d system, a characteristic saddle-node bifurcation results as endpoint of the branch of linearly stable pinned droplets (III), which occurs at a critical drop volume $V_\mathrm{D} \approx 3.19\times10^4$. The deformation of the contact line by the heterogeneity is clearly visible in the height profile of this critically adhering drop. 
Analogous to the 1d solutions in Fig.~\ref{fig:dynhe_volcont_1D} also in Fig.~\ref{fig:dynhe_volcont_2D} branches of unstable solutions (IV) occur that are connected by a cascade of saddle-node bifurcations.

\subsection{Depinning-pinning cycles} 
\label{sec:supp-slip}
\begin{figure} \center
\includegraphics[width=0.7\hsize]{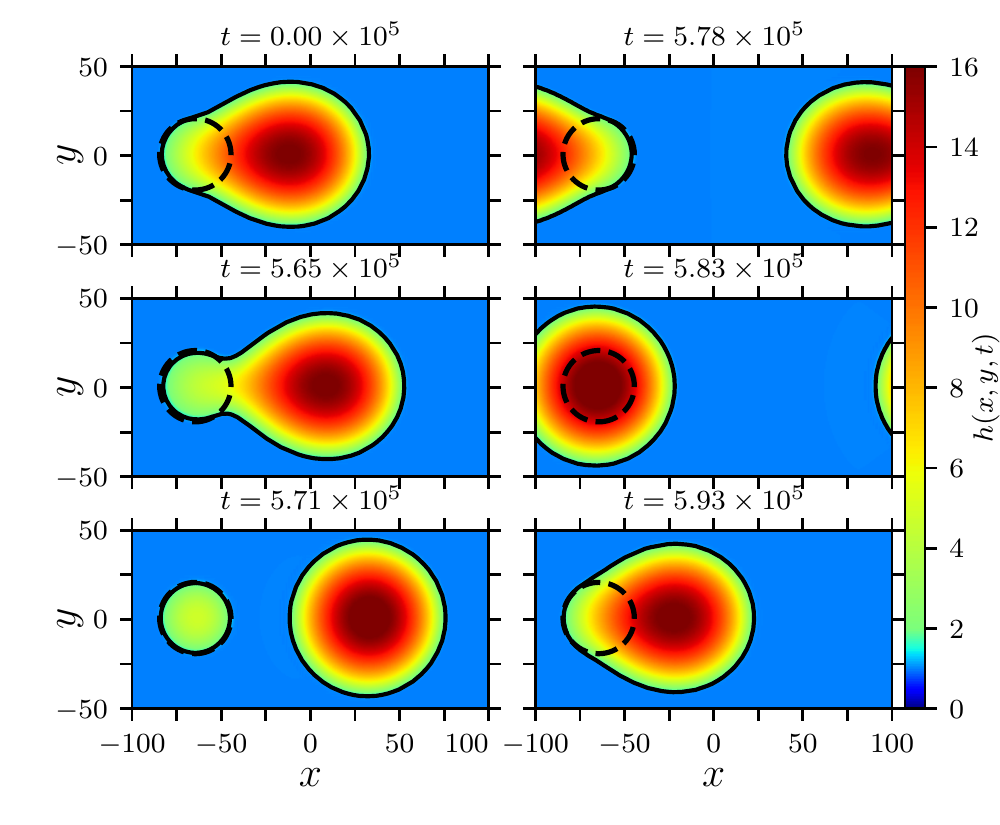}
\caption{Snapshots from the dynamics of a depinned drop beyond the critical slope, i.e., at $\alpha=0.1934 \gtrsim \alpha_\mathrm{depin}$. After a relatively long phase where a nonstationary drop is pinned (until $t \gtrsim 5.65\times 10^5$), the main droplet depins from the hydrophilic spot. After the drop has passed through the periodic boundary it collides again with the hydrophilic spot at $t \approx 5.78\times10^5$ and pins quasi-stationarily to it, resulting in an overall time-periodic stick-slip cycle. The drop volume is $V_\mathrm{D} = 5\times10^4$ and the remaining parameters are as in Fig.~\ref{fig:dynhe_volcont_2D}.}
\label{fig:dynhe_depin_2D}
\end{figure}

\begin{figure} \center
\includegraphics[width=0.7\hsize]{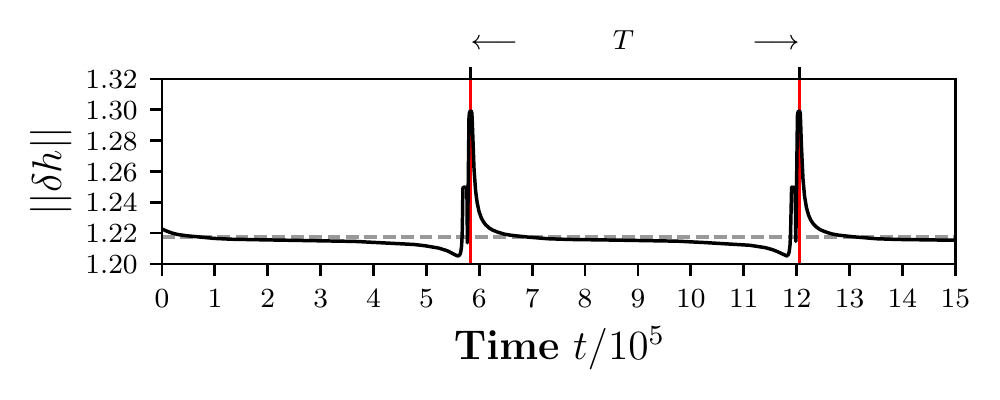}\\
\includegraphics[width=0.7\hsize]{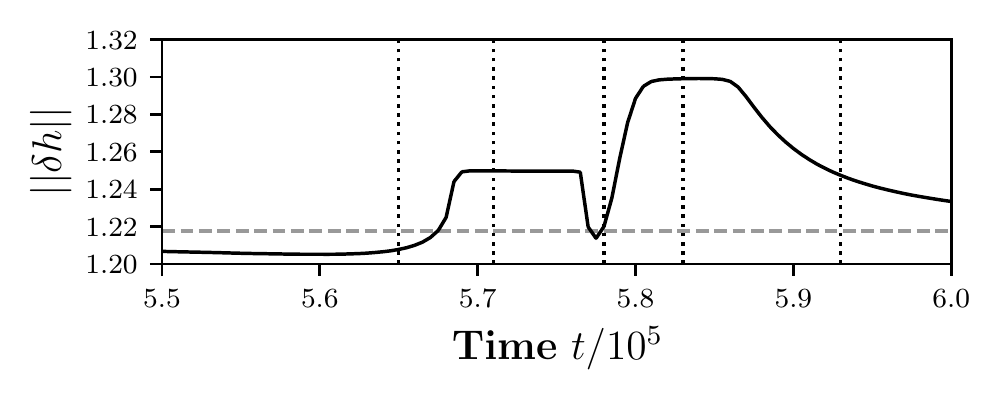}
\caption{The top panel shows the $L^2$-norm $||\delta h||$ over time $t$ for a drop with volume $V_\mathrm{D} = 5\times10^4$ at fixed $\alpha=0.1934 \gtrsim \alpha_\mathrm{depin}$ at parameters as in Fig.~\ref{fig:dynhe_depin_2D}. The part between the two peaks indicated by arrows represents one stick-slip cycle. The separation of the stick and slip phases is illustrated here by the length of the central plateau as compared to the characteristic peak width. Based on the maxima, a period $T\approx 6.21 \times 10^5$ is be determined. In addition, the time average $\overline{||\delta h||} \approx 1.218$ is shown as horizontal dashed line. The lower panel focuses onto the first peak and indicates with vertical dotted lines the times for which Fig.~\ref{fig:dynhe_depin_2D} gives profiles.}
\label{fig:dynhe_depin_L2_time}
\end{figure}

The dynamics near the saddle-node bifurcation at $\alpha_\mathrm{depin} \approx 0.1926$ (see Fig.~1 of main text) is next studied by time simulations using the critical pinned drop as initial condition (IC) at slightly increased inclination. We start with the critical pinned steady solution of the continuation result, increase $\alpha$ to $0.1934 \gtrsim \alpha_\mathrm{depin}$ and obtain the dynamics presented in the form of profile snapshots in Fig.~\ref{fig:dynhe_depin_2D}. The corresponding time dependence of the norm is given in Fig.~\ref{fig:dynhe_depin_L2_time}. The original steady drop, which is already deformed by the interplay of heterogeneity and lateral force, further elongates  in the direction of the driving force on a comparatively large time scale. 

The liquid bridge between the drop and the hydrophilic spot eventually breaks at $t \approx 5.65\times 10^5$, i.e., after the stretching phase depinning and slipping is observed. On the hydrophilic spot a thick liquid film remains. At $t = 5.71\times10^5$ the drop slides on a locally homogeneous substrate and has an almost spherical cap-like shape due to the relatively low inclination (see Fig.~\ref{fig:dynhe_depin_2D}). 
Since here we use periodic boundary conditions in $x$-direction, the main drop, after passing through the boundary at $t \approx 5.78\times10^5$, reaches the hydrophilic spot again, covers it and remains for quite some time nearly pinned (but again slowly stretching).

In total, this results in a stick-slip cycle with a clear separation of the time scales by an order of magnitude: For a time intervall of $t_\mathrm{pin} \approx 6\times 10^5$ the drop -- although not as a steady state -- pins to the heterogeneity. After the drop depins, it slides for a time intervall of $t_\mathrm{slide} \approx 4\times 10^4$ on the locally homogeneous substrate and then returns close to the pinned state. If one calculates the temporal mean value $\overline{||\delta h||}$ of these solutions for different inclinations, one obtains the time-periodic branch shown in Fig.~1 of the main text. 

\begin{figure} \center
\includegraphics[width=0.8\hsize]{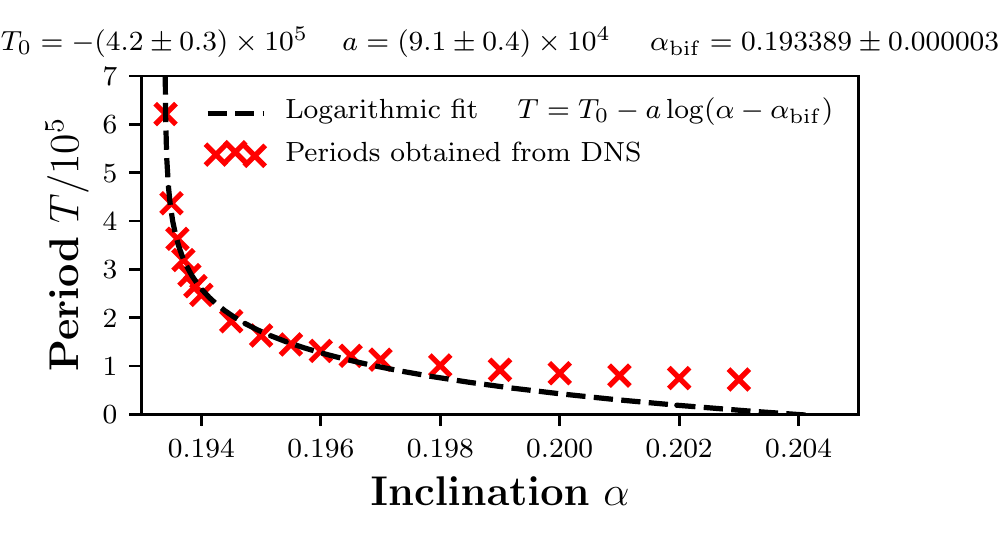}
\caption{Shown is the dependency of the period of the stick-slip cycle on the inclination angle $\alpha>\alpha_\mathrm{bif}$ as obtained by DNS (cross symbols). A logarithmic fit is given as dashed line and indicates the occurrence of a homoclinic bifurcation. It implies $T \rightarrow \infty$ for $\alpha \searrow \alpha_\mathrm{bif} \approx 0.19339$. At large inclination one finds a deviation from the logarithmic behaviour. Remaining parameters are as in Fig.~\ref{fig:dynhe_depin_2D}.}
\label{fig:dynhe_depin_period}
\end{figure}

Analogously to the pearling-coalescence cycle analysed in Ref.~\cite{EWGT2016prf}, the period $T$ depends significantly on the streamwise domain length $l_x$. However, this dependency exclusively results from the length of the sliding phase $t_\mathrm{slide}$. Close to the depinning bifurcation, it is always much shorter than the length of the depinning and pinning processes, i.e., the overall period is largely independent of the domain length.

The averaged $||\delta h||$ rises significantly with $\alpha$ in the range close to the bifurcation, but remains almost constant at larger inclination angles $\alpha$ (Fig.~1 of main text). Fig.~\ref{fig:dynhe_depin_L2_time} shows the dynamics of $||\delta h||$ during individual cycles, that reflects changes of the thickness profile. At depinning it increases, then it is nearly constant during the short sliding phase (see lower panel of Fig.~\ref{fig:dynhe_depin_L2_time}), first decreases then sharply increases at pinning before showing another short plateau when the spot is fully covered by the again cap-like drop. Then the long stretching phase follows marked by a slow decrease of the norm. Then the cycle restarts.

We expect that for significantly larger angles, comparable to $\alpha_\mathrm{pearl}$, the pinning/depinning cycle shows also the pearling instability and the dynamics becomes more involved (cf.~period doubling route to chaos in Ref.~\cite{EWGT2016prf}). Here, this is not considered further. The cycle period $T$ defined in Fig.~\ref{fig:dynhe_depin_L2_time} is analysed quantitatively for different inclinations and presented in Fig.~\ref{fig:dynhe_depin_period}. Close to the bifurcation $T$  can be very well fitted by a logarithm. This indicates that the underlying global bifurcation is a homoclinic bifurcation \cite{Strogatz2014}. However, this seems to contradict Fig.~1 of the main text as the branch of time-periodic states seems to emerge directly from the saddle-node bifurcation. This indicates, in contrast, that the global bifurcation is a SNIPER bifurcation (short for Saddle-Node Infinite PERiodic Bifurcation) \cite{Kuznetsov2010,Strogatz2014} as found for other depinning transitions \cite{ThKn2006prl,ThKn2006njp}. The situation does not change when zooming in onto the bifurcation point. The issue cannot be unambiguously resolved within the framework of the available numerical resolution. There is a very small but significant deviation between the $\alpha_\mathrm{depin} \approx 0.19257$ obtained in the numerical continuation and $\alpha_\mathrm{bif} \approx 0.19339$ as obtained in the time simulations. The difference corresponds to a relative error of $0.4$\% and indicates a limited comparability of the two different numerical methods. However, the characteristic dependency for the SNIPER bifurcation $T \sim (\alpha-\alpha_\mathrm{bif})^{-1/2}$ could be clearly excluded with the data in Fig.~\ref{fig:dynhe_depin_period}. We therefore conclude that the global bifurcation is indeed a homoclinic bifurcation that most likely occurs at slightly smaller $\alpha$ than the saddle-node bifurcation.

\subsection{Power laws for depinning threshold} 
\label{sec:supp-power}
\begin{figure} \center
\includegraphics[width=0.8\hsize]{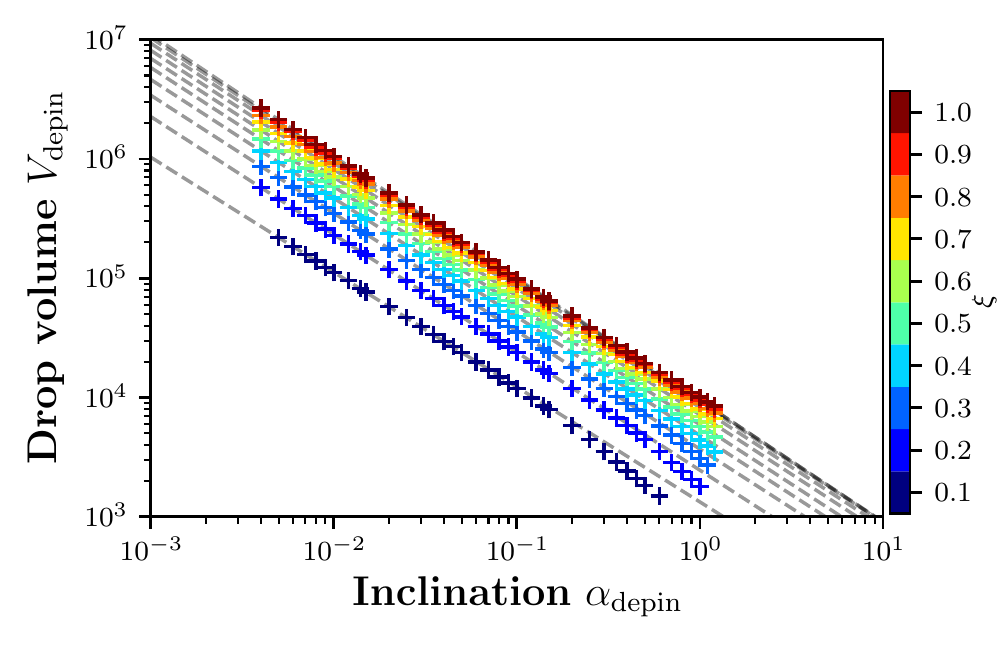}
\caption{Power laws $V_\mathrm{D}(\alpha_\mathrm{depin})$ for the depinning threshold for various amplitudes $\xi$ of the wettability contrast. Qualitatively the dependency does not change with $\xi$ and can always be fitted by $V_\mathrm{depin} = a_{\xi}\alpha_\mathrm{depin}^{b_{\xi}}$. The power $b_\xi$ deviates from $b_\xi \approx -1.0$ only in the range of small and large $\xi$. Increasing $\xi$ mainly causes a parallel shift in the log-log representation, i.e., manly the coefficient $a_\xi$ changes. The numerical values are given in Table~\ref{tab:power}.}
\label{fig:dynhe_depin_powerlaw_all}
\end{figure}

To quantify the results found for different drop volumes, the bifurcation diagram of the steady drops is calculated for many different values of $V_\mathrm{D}$ and the wettability contrast $\xi$. The loci of the saddle-node bifurcations where the depinning occurs are summarised in the parameter plane spanned by $V_\mathrm{D}$ and $\alpha$ (see Fig.~\ref{fig:dynhe_depin_powerlaw_all}). If the drop volume is increased, the critical slope $\alpha_\mathrm{depin}$ shifts to smaller values following a power law. An increase in the wettability contrast $\xi$ shifts the power law towards larger drop volumes. The diameter of the hydrophilic spot has no significant influence on the power laws shown in Fig.~\ref{fig:dynhe_depin_powerlaw_all}.

\begin{table*} \center
%{\scriptsize
{\small
\begin{tabular}{| c || c | c | c | c | c | c | c | c | c | c |} \hline
$\xi$ & $0.1$ & $0.2$ & $0.3$ & $0.4$ & $0.5$ & $0.6$ & $0.7$ & $0.8$ & $0.9$ & $1.0$ \\\hline\hline
$a_{\xi}/10^3$ & $1.33$ & $2.46$ & $3.68$ & $4.81$ & $5.92$ & $7.16$ & $8.23$ & $9.07$ & $9.40$ & $9.48$ \\\hline
$b_{\xi}$ & $-0.96$ & $-0.99$ & $-0.99$ & $-0.99$ & $-1.00$ & $-1.00$ & $-1.00$ & $-1.00$ & $-1.01$ & $-1.02$ \\\hline
\end{tabular}
}
\caption{Numerical parameters for the power law fits of the curves in Fig.~\ref{fig:dynhe_depin_powerlaw_all} for various values of the wettability contrast $\xi$.}
\label{tab:power}
\end{table*}

\begin{figure} \center
\includegraphics[width=0.7\hsize]{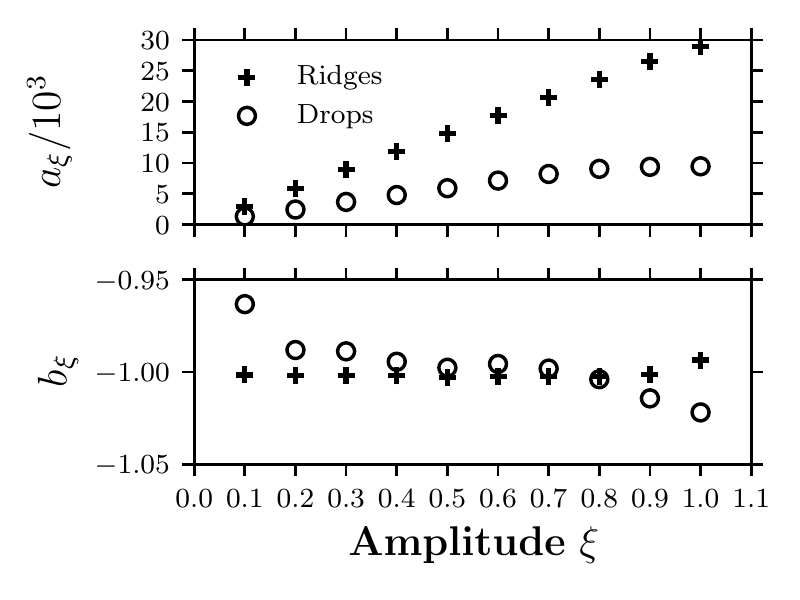}
\caption{Shown is the dependency of the coefficients of the power law fits of the curves in Fig.~\ref{fig:dynhe_depin_powerlaw_all} on the wettability contrast $\xi$. Results are shown for 1d drops (ridges) and 2d drops. In both cases, the power remains quite close to $b_\xi \approx -1.0$ and the coefficient $a_\xi$ shows a linear increase with increasing $\xi$. For the 2d drops there is a more pronounced deviation from this behaviour in the range of small and large values of $\xi$.}
\label{fig:dynhe_depin_powerlaw_coeffs}
\end{figure}

The dependency of the coefficients of the power law fits of the curves in Fig.~\ref{fig:dynhe_depin_powerlaw_all} on the wettability contrast $\xi$ is given in Fig.~\ref{fig:dynhe_depin_powerlaw_coeffs} together with results for 1d drops.  In both cases -- 1d and 2d drops -- the power law remains nearly linear and the coefficient $a_\xi$ shows a linear increase with $\xi$. Only for $\xi < 0.3$ there is a deviation in the range of small drops, which are of secondary importance for the underlying physical system and are therefore excluded from the fit. In the double-logarithmic representation of Fig.~\ref{fig:dynhe_depin_powerlaw_all}, the fit functions are nearly parallel, so that the power $b_\xi$ appears essentially independent of the wettability contrast. The parallel shift towards larger drop volumes for increasing $\xi$ corresponds to the expected behaviour and is linked to the coefficient $a_\xi$. 

\begin{figure}[!b] \center
\includegraphics[width=0.8\hsize]{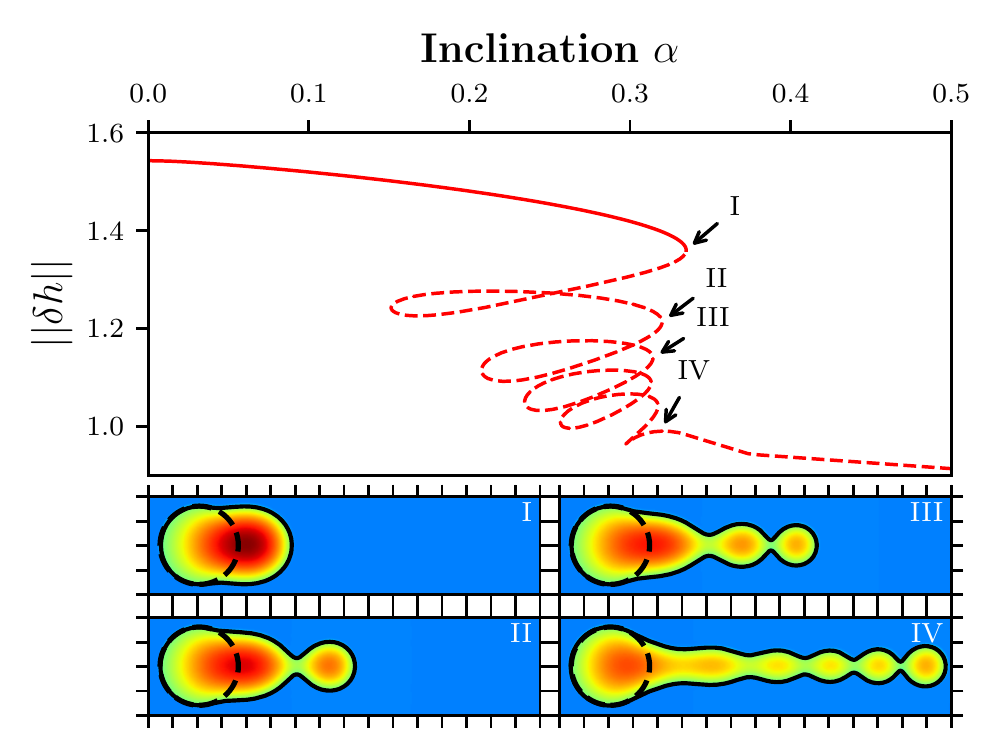}
\caption{(top) Bifurcation diagram and (bottom) height profiles of selected elongated pinned drops at loci indicated by the roman numbers.
The norm is shown as a function of inclination $\alpha$ for fixed drop volume $V_\mathrm{D} = 6.2\times10^4$. The domain is enlarged in $x$-direction: $l_x = 4l_y = 400$ and features a hydrophilic spot of diameter $2R = 80$ with wettability contrast $\xi = 1.0$.} 
\label{fig:dynhe_depin_longdrops}
\end{figure}

\subsection{Elongated pinned drops} 
\label{sec:supp-elongated}
As already indicated by the numerical continuations in the volume (see Fig.~\ref{fig:dynhe_volcont_2D}) on a relatively small domain with $l_x = 2l_y = 200$, one may find elongated steady profiles of pinned drops with downstream satellite drops. To emphasise essential aspects of these interesting states, we discuss the bifurcation diagram discussed obtained for a larger domain $l_x = 4l_y = 400$ (longer compared to the one used for Fig.~\ref{fig:dynhe_volcont_2D}) and a larger hydrophilic spot of radius $R = 40$. However, the multi-drop solutions are all linearly unstable (for most of them more than one eigenvalue has a positive real part). Therefore, we expect in contrast to the elongated sliding drops in \cite{EWGT2016prf} that they only play a minor role in the statistics of drop ensembles. In consequence, we only consider them in passing.

Fig.~\ref{fig:dynhe_depin_longdrops} gives the bifurcation diagram and height profiles of selected elongated pinned drops. At  $\alpha = 0.0$, the cap-like drop is slightly larger than the hydrophilic spot that it symmetrically covers. When the substrate is inclined, the drop deforms such that the $L_2$-norm monotonically decreases. As above for the smaller domain, a first characteristic saddle-node bifurcation occurs (here at $\alpha_\mathrm{depin} \approx 0.3350$), which determines the transition from stable pinned drops to depinned sliding drops that undergo a stick-slip motion. %The transition is captured by the power laws as presented above.

At $\alpha_\mathrm{depin}$, the branch of steady pinned drops folds back and subsequently undergoes an intricate sequence of ten saddle-node bifurcations. At these bifurcations the pinned solutions become more and more unstable. The linearly unstable solutions correspond to a longer and longer chain of smaller drops, which are interconnected by narrower neck-like regions. Every second saddle-node bifurcation corresponds to the addition of another drop to the chain. For example, the solution (II) at the third bifurcation has one additional drop and the solution (III) at the fifth bifurcation has two additional drops. For the used domain length, a total of five additional drops is possible (profile IV) before the downstream tip interacts with the drop on the hydrophilic spot due to the periodic boundary conditions. The solution then becomes a modulated rivulet aligned in $x$-direction very similar to states found for inclined substrates with a stripe-like heterogeneity, e.g., solution SD$_3$ at large $\mu$ in Figs.~13(a) and 14(c) of \cite{BKHT2011pre}.

\begin{figure}[!b] \center
\includegraphics[width=0.7\hsize]{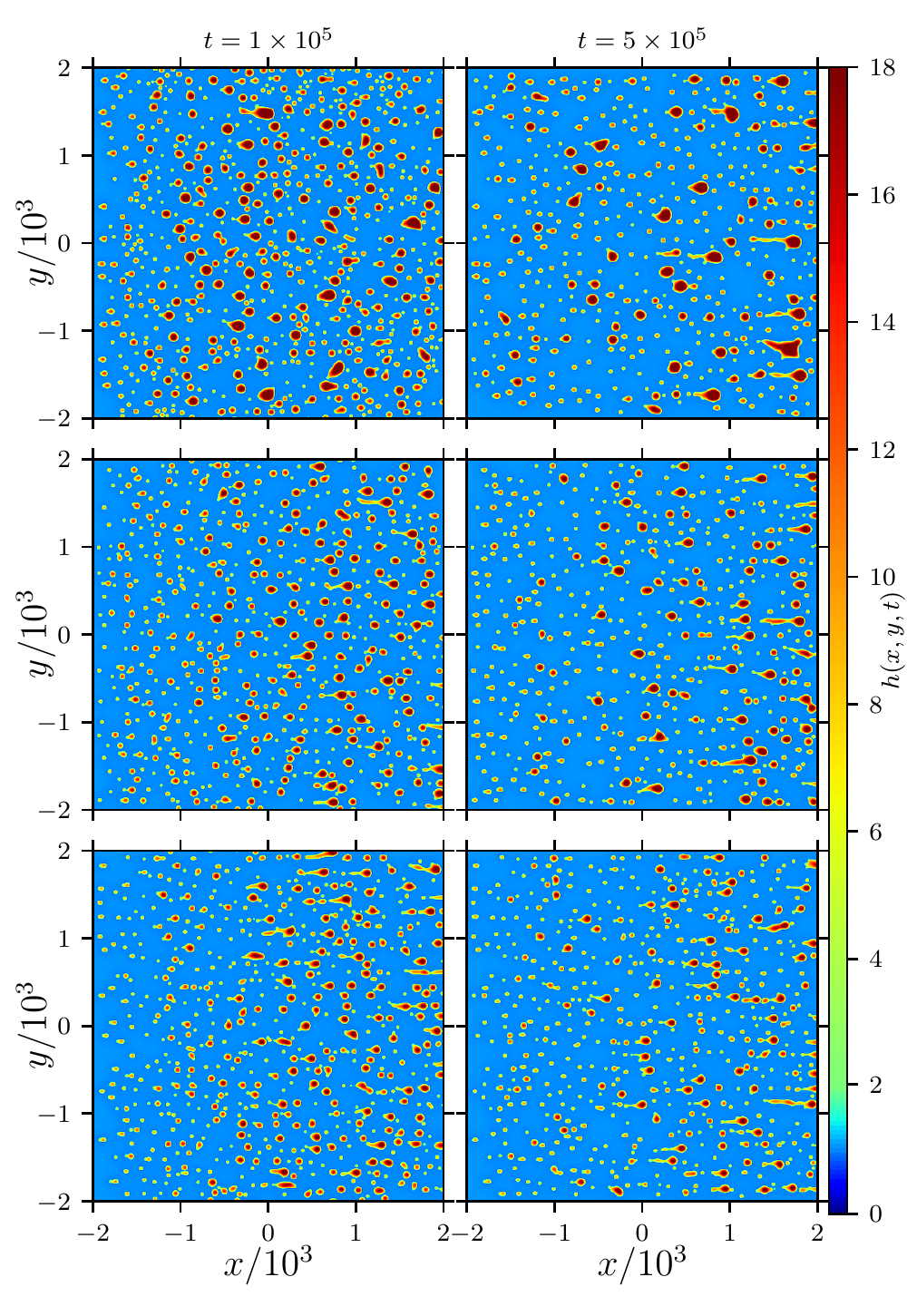}
\caption{Snapshots from a time evolutions of a drop ensemble for (top) $\alpha = 0.3$, (middle) $\alpha = 0.5$, and (bottom) $\alpha = 0.7$ with $N_\mathrm{S} \approx 400$ randomly distributed hydrophilic spots and condensation rate $\beta_\mathrm{KE} = 1\times10^{-4}$. In each case one snapshot before and one after convergence to a stationary drop size distribution is shown.}
\label{fig:dynhe_ense_tseries_smallalp}
\end{figure}

We would therefore expect the solution to gain stability at an inclination far outside the range of Fig.~\ref{fig:dynhe_depin_longdrops} as discussed at Fig.~25 of \cite{BKHT2011pre}. Note that overall, profiles and bifurcation diagrams found here show clear similarities to results obtained with a model for the dripping faucet \cite{CoMR2000ptps,CoMR2005jfm}. This observation makes it likely, that also here an extended parameter study would reveal the full period-doubling route to chaos as for the pearling instability of sliding drops \cite{WTEG2017prl}.

\subsection{Ensemble behaviour}
\label{sec:supp-ensemble}
In addition to the case shown in the main text (moderate evaporation rate $\beta = 10^{-4}$ and inclination $\alpha = 0.5$) here we present further images of drop ensembles at all inclination angles employed employed in the ensemble DNS in the main text. Fig.~\ref{fig:dynhe_ense_tseries_smallalp} shows snapshots at identical $\beta$ at inclinations $\alpha = 0.3, 0.5$ and $0.7$. In each case, the snapshots at $t = 1.0\times 10^5$ gives an impression of the transient phase where the drop size distribution is still changing, while the snapshots at $t = 5.0\times 10^5$ represent already the emerged statistically stationary state. For all $\alpha$ the stationary state is characterised by many relatively small pinned drops and larger sliding drops. With decreasing $\alpha$ one finds significantly fewer sliding drops that have, however, larger volumes. In general, the lower the inclination $\alpha$, the larger the volume $V_\mathrm{depin}$ at the depinning threshold, and the ensemble consists of fewer larger drops because at lower $\alpha$ depinning occurs later. As the depinned drops also slide more slowly the mean height of liquid in the domain is larger at lower $\alpha$ (cf.~Fig.~5 of main text).

Note, that in consequence the footprint of the larger drops at small $\alpha$ often show extended backwards protrusions and the drops extend over a length larger than the mean distance of the hydrophilic spots. Therefore, usually several spots are covered resulting in drop deformations. It would be of interest to investigate in the future the distribution of the drops with protrusion for different inclinations and to relate it to the detailed structure of unstable branches in single-drop bifurcation diagrams as shown in Fig.~\ref{fig:dynhe_depin_longdrops}.

\end{document}